\documentclass[twocolumn,aps,showpacs,nofootinbib,prd]{revtex4-1}
\usepackage{amssymb}
\usepackage{amsmath}
\usepackage{graphics}
\usepackage{epsfig}
\usepackage[dvips]{color}
\newcommand{\be}{\begin{equation}}
\newcommand{\ee}{\end{equation}}
\newcommand{\bea}{\begin{eqnarray}}
\newcommand{\eea}{\end{eqnarray}}
\newcommand{\beas}{\begin{eqnarray*}}
\newcommand{\eeas}{\end{eqnarray*}}

\newcommand{\nn}{\nonumber\\}
\newcommand{\slsh}[1]{{\not \! #1}}

\begin{document}
\title{Thermal photon production from gluon fusion induced by magnetic fields in relativistic heavy-ion collisions}
\author{Alejandro Ayala$^{1,2}$, Jorge David Casta\~no-Yepes$^1$, C. A. Dominguez$^2$, L. A. Hern\'andez$^2$}
  \address{
  $^1$Instituto de Ciencias
  Nucleares, Universidad Nacional Aut\'onoma de M\'exico, Apartado
  Postal 70-543, M\'exico Distrito Federal 04510,
  Mexico.\\
  $^2$Centre for Theoretical and Mathematical Physics, and Department of Physics,
  University of Cape Town, Rondebosch 7700, South Africa}

\begin{abstract}
We compute the production of thermal photons in relativistic heavy-ion collisions by gluon fusion in the presence of an intense magnetic field, and during the early stages of the reaction. This photon yield is an excess over calculations that do not consider magnetic field effects. We add this excess to recent hydrodynamic calculations that are close to describing the experimental transverse momentum distribution in RHIC and LHC.  We then show that with reasonable values for the temperature,  magnetic field strength, and strong coupling constant, our results provide a very good description of such excess. These results support the idea that the origin of at least some of the photon excess observed in heavy-ion experiments may arise from magnetic field induced processes.

\end{abstract}

\maketitle

\begin{widetext}
\begin{figure*}
\begin{center}
\includegraphics[scale=1]{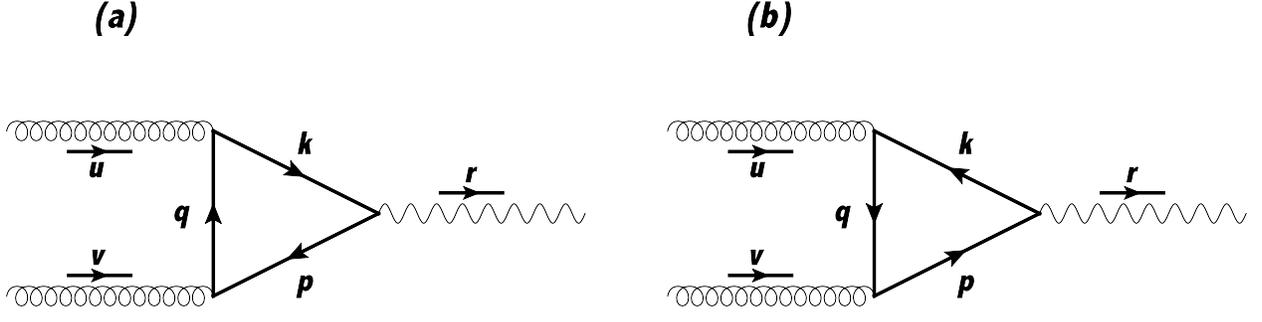}
\end{center}
\caption{Feynman diagrams for the amplitude for photon production from gluon fusion. The thick lines in the loop represent the quark propagators in the presence of the magnetic field.}
\label{fig1}
\end{figure*}
\end{widetext}

The results from heavy-ion experiments carried out at the BNL Relativistic Heavy-Ion Collider (RHIC), and at the CERN Large Hadron Collider (LHC), show that a state of matter is formed where quarks and gluons are not confined to individual nucleons~\cite{RHIC, LHC}. Non-central collisions  produce magnetic fields with intensity at the beginning of the reaction is estimated to be as high as several times the mass of the pion squared~\cite{intensity1,intensity2, intensity3}.  Although the intensity of these fields fades out fast with time, their strength lasts long enough after the onset of the collision to induce observable phenomena, for instance, the charge separation along the field direction~\cite{chargesep}, which is linked to the chiral magnetic effect~\cite{intensity1,Fukushima}.\\ 
A magnetic field makes it possible to produce photons from processes otherwise not allowed. For instance, it has been shown that the QCD trace anomaly can turn the energy momentum of the soft gluon bulk into photons~\cite{Skokov, Shuryak}. In addition, quarks can emit photons by synchrotron radiation~\cite{Tuchin}. Other approaches to study photon production in the presence of an intense magnetic field include the gauge/gravity correspondence in a strongly coupled ${\mathcal{N}}= 4$ plasma~\cite{Leonardo}. These novel calculations have recently been implemented  to explain the experimentally measured excess~\cite{experiments} of thermal photons over models that describe well other low momentum observables. The enhanced production of photons in heavy-ion reactions has also been studied in the absence of magnetic field effects, {\it e.g.} by assuming the modification of the quark and gluon distributions to be a power-like tail at high energies~\cite{McLerran}, or by the delayed formation of the quark-gluon plasma~\cite{Liu}.\\
A magnetic field naturally produces an asymmetry in the emission of electromagnetic radiation. Therefore, magnetic fields can also be a source of not only an excess in the photon yield, but also of the puzzling large strength of the coefficient $v_2$ in the Fourier expansion of the azimuthal distribution. The latter has been found to be as large as that of pions~\cite{v2photons}. Although some recently improved hydrodynamic \cite{hydro-photons1,hydro-photons2} and transport~\cite{transport} calculations obtain a better agreement with ALICE and PHENIX measurements of low transverse momentum photons, this agreement is not yet complete~\cite{review}. Therefore, it remains important to quantify the fraction of the yield, and of the asymmetry arising from magnetic field effects, to better characterize the initial stages of heavy-ion reactions.\\ 
In non-central collisions, and at early times, the magnetic field is of high intensity, the largest temperatures are achieved, and the soft dynamics is dominated by gluons~\cite{Shuryak}. It is then natural to explore a mechanism where collisions of these gluons induce the emission of photons. In this work we compute, and to our knowledge for the first time, the production of thermal photons from the perturbative fusion of gluons at these early collision times.\\ 
The amplitude for the process is depicted by the Feynman diagrams in Fig.~\ref{fig1}, which also defines the kinematical variables. The thick loop lines represent the quark propagator in the presence of the magnetic field. In the absence of this field, the diagrams cancel each other. It is the presence of the field which makes it possible that both diagrams contribute with the same relative sign. \\

The fermion propagator in coordinate space cannot longer be written as a simple Fourier transform of a momentum propagator but instead it is written as~\cite{Schwinger}
\bea
   S(x,x')=\Phi (x,x')\int\frac{d^4p}{(2\pi)^4}e^{-ip\cdot (x-x')}S(p)\;,
\label{genprop}
\eea
where
\bea
   \!\!\!\Phi (x,x')=\exp\left\{iq_f\int_{x'}^xd\xi^\mu\left[A_\mu + \frac{1}{2}F_{\mu\nu}(\xi - x')^\nu\right]\right\}\!\!,
\label{phase}
\eea
is called the {\it phase factor}, and $q_f$ is the absolute value of the quark charge. We consider the contribution of two light flavors, thus $q_u = 2|e|/3$ and $q_d = |e|/3$. The propagator in momentum-space, $S(p)$, is given by
\begin{eqnarray}
   &&  \!\!\!\! i S(p) = \int _{0}^{\infty}\frac{ds}{\cos(q_fBs)}
   e^{is(p_{\parallel}^{2} - p_{\perp}^{2}\frac{\tan (q_fBs)}{q_fBs}- m_f^{2} +i\epsilon)}\nonumber\\
   && \!\!\!\!\biggl[\!\left(\cos (q_fBs)\!\! + \!\! \gamma_1 \gamma_2 \sin (q_fBs)\right)
   \!\!(m_f+\slsh{p_{\|}})\!\!  - \!\!  \frac{\slsh{p_\bot}}{\cos(q_fBs)} \biggr]\! ,
   \label{tracewithSchwinger}
 \end{eqnarray}
where $m_f$ is the quark mass. We have chosen the homogeneous magnetic field to point in the $\hat{z}$ direction, namely $\boldsymbol{B}=B\hat{z}$. This configuration can be obtained from an external vector potential which we choose in the so called {\it symmetric gauge} $A^{\mu}= \frac{B}{2}(0,-y,x,0)$. We have also defined $
p_\perp^\mu\equiv(0,p_1,p_2,0),\
p_\parallel^\mu\equiv(p_0,0,0,p_3),\
p_\perp^2\equiv p_1^2+p_2^2$ and
$p_\parallel^2\equiv p_0^2-p_3^2$.

Since the two Feynman diagrams of Fig.~\ref{fig1} give the same contribution, we concentrate on the computation of the amplitude depicted in Fig.~\ref{fig1}a which becomes
\bea
   {\mathcal{M}}^{(a)}&=&-\int \!d^4x\int \!d^4y\int \!d^4z\int\!\frac{d^4p}{(2\pi)^4}
   \int\!\frac{d^4q}{(2\pi)^4}\int\!\frac{d^4k}{(2\pi)^4}
   \nonumber\\
   &\times&e^{-ip\cdot (y-x)}e^{-iq\cdot (z-y)}e^{-ik\cdot (x-z)}e^{-iu\cdot z}e^{-iv\cdot y}e^{ir\cdot x}\nonumber\\
   &\times&
   {\mbox{Tr}}\left[ iq_f\gamma_\mu iS(k) ig\gamma_\alpha t^c iS(q) ig\gamma_\nu t^d iS(p) \right]
   \epsilon^{*\mu}(\lambda_r)\nonumber\\
   &\times&\Phi(x,y)\Phi(y,z)\Phi(z,x)\epsilon^\alpha(\lambda_u)\epsilon^\nu(\lambda_v).
   \label{amplitude}
\eea
The product of phase factors can be written as
\bea
   \Phi(x,y)\Phi(y,z)\Phi(z,x)
   =e^{i\frac{q_fB}{2}\epsilon_{ij}(z-x)_i(x-y)_j},
   \label{prodphase}
\eea
where we used the explicit form of $A^\mu$ which gives $F_{12}=-F_{21}=-B$, with the rest of the components of $F_{\mu\nu}$ vanishing, and $\epsilon_{ij}$ being the Levi-Civita symbol. The integrations in Eq.~(\ref{amplitude}) are carried out more easily by making the change of variables $\omega=z-x$ and $l=x-y$, after which the integration over the spatial coordinates becomes
\bea
   &&(2\pi)^4\delta^4(r-v-u)\int d^4\omega\int d^4l\
   e^{-i\omega\cdot (q-k+u)}e^{-il\cdot (q-p-v)}\nonumber\\
   &&e^{i\frac{q_fB}{2}\epsilon_{ij}\omega_i l_j},
   \label{afterchange}
\eea
which exhibits the overall energy-momentum conservation in the process. Furthermore, we split the integrals over $\omega$ and $l$ in transverse and longitudinal parts
\bea
   &&\int d^4\omega\int d^4l\
   e^{-i\omega\cdot (q-k+u)}e^{-il\cdot (q-p-v)}e^{i\frac{q_fB}{2}\epsilon_{ij}\omega_i l_j}\nonumber\\
   =&&(2\pi)^4\delta^2\left[(q-k+u)_\parallel\right]
   \delta^2\left[(q-p-v)_\parallel \right]\nonumber\\
   \times&&\prod_{i,j=1,2}\int d\omega_i\int dl_j e^{i\omega_i(q-k+u)_i}e^{il_j(q-p-v)_j}
   e^{i\frac{q_fB}{2}\epsilon_{ij}\omega_i l_j}\nonumber\\
   =&&(2\pi)^4\delta^2\left[(q-k+u)_\parallel\right]
   \delta^2\left[(q-p-v)_\parallel \right]\nonumber\\
   \times&&\left(\frac{4\pi}{q_fB}\right)^2\prod_{i,j=1,2}
   e^{i\frac{2}{q_fB}\epsilon_{ij}(q-k+u)_i(q-p-v)_j}.
   \label{separate}
\eea
This shows that the longitudinal momentum is explicitly conserved at the vertices, whereas the transverse momentum is not but instead its components are mixed up by the magnetic field.\\
We now use the fact that  the quark dynamics is dominated by the lowest Landau level when  the magnetic field is very intense. For this level, the propagator in Eq.~(\ref{tracewithSchwinger}) can explicitly be written as
\bea
   iS(p)=2ie^{-\frac{p_\perp^2}{q_fB}}\frac{(\slsh{p_{\parallel}}+m_f)}{p_\parallel^2-m_f^2}
   \left[ \frac{1- i\gamma_1\gamma_2}{2} \right].
   \label{propLLL}
\eea
The operator ${\mathcal{O}}_\parallel=\left[1- i\gamma_1\gamma_2 \right]/2$ projects onto the longitudinal space. Therefore the matrix element can be factorized into a product of transverse and longitudinal pieces, namely
\bea
   {\mathcal{M}}^{(a)}&=&(2\pi)^4\delta^4(r-v-u){\mathcal{M}}^{(a)}_\perp{\mathcal{M}}^{(a)}_\parallel
   \label{factorized}\\
   {\mathcal{M}}^{(a)}_\perp&=&\left(\frac{4\pi}{q_fB}\right)^2\int\frac{d^2p_\perp}{(2\pi)^2}
   \int\frac{d^2q_\perp}{(2\pi)^2}\int\frac{d^2k_\perp}{(2\pi)^2}\nonumber\\
   &\times&
   e^{-\!\frac{k_\perp^2}{q_fB}}e^{-\!\frac{q_\perp^2}{q_fB}}e^{-\!\frac{p_\perp^2}{q_fB}}
   \!\!\!\!\prod_{i,j=1,2}e^{i\frac{2}{q_fB}\epsilon_{ij}(q-k+u)_i(q-p-v)_j}\nonumber\\
   &=&\left(\frac{q_fB}{12\pi}\right)e^{-\frac{(u+v)_\perp^2}{3q_fB}},
   \label{factorperp}\\
   {\mathcal{M}}^{(a)}_\parallel&=&-8\left(\frac{q_fg^2\delta^{cd}}{2}\right)\int\frac{d^2p_\parallel}{(2\pi)^2}
   \int\frac{d^2q_\parallel}{(2\pi)^2}
   \int\frac{d^2k_\parallel}{(2\pi)^2}\nonumber\\
   &\times&(2\pi)^4\delta^2\left[(q-k+u)_\parallel\right]
   \delta^2\left[(q-p-v)_\parallel \right]\epsilon^{*\mu}(\lambda_r)\nonumber\\
   &\times&\frac{{\mbox{Tr}}\left[\gamma_\mu\slsh{k_{\parallel}}{\mathcal{O}}_\parallel 
   \gamma_\alpha\slsh{q_{\parallel}}{\mathcal{O}}_\parallel 
   \gamma_\nu\slsh{p_{\parallel}}{\mathcal{O}}_\parallel\right]}
   {k_\parallel^2q_\parallel^2p_\parallel^2}
   \epsilon^\alpha(\lambda_u)\epsilon^\nu(\lambda_v).
   \label{factorpara}
\eea
Since at the early stages of the collision gluons are far more abundant than quarks, we compute Eq.~(\ref{factorpara}) under the assumption that quarks do not yet thermalize. Accordingly, we set $m_f=0$ since in the absence of thermal corrections, the light-quark vacuum masses are negligible. The trace in Eq.~(\ref{factorpara}) contains the product of up to twelve gamma matrices. The resulting expression is long and involved. It is however easy to show that upon squaring and summing over polarizations, only a small piece survives so that the trace can be expressed as
\bea
   &&{\mbox{Tr}}\left[\gamma_\mu\slsh{k_{\parallel}}{\mathcal{O}}_\parallel 
   \gamma_\alpha\slsh{q_{\parallel}}{\mathcal{O}}_\parallel 
   \gamma_\nu\slsh{p_{\parallel}}{\mathcal{O}}_\parallel\right]
   \longrightarrow
   k_{\parallel\nu}(p_{\parallel\mu}q_{\parallel\alpha} - p_{\parallel\alpha}q_{\parallel\mu})\nn
   &&+
   k_{\parallel\mu}(p_{\parallel\nu}q_{\parallel\alpha} + p_{\parallel\alpha}q_{\parallel\nu})
   +
   k_{\parallel\alpha}(p_{\parallel\nu}q_{\parallel\mu} - p_{\parallel\mu}q_{\parallel\nu}),
   \label{tracesur}
\eea
where the arrow indicates this to be the only contributing portion. Two of the integrations in Eq.~(\ref{factorpara}) become straightforward    using the delta-function restrictions. We choose those two as the integrals over $k_\parallel$ and $q_\parallel$. The remaining integral contains the product of momenta in the denominator that we write in its Feynman parametrization form
\begin{widetext}
\bea
   \frac{1}{p_\parallel^2 (p+v)_\parallel^2 (p+v+u)_\parallel^2}
   &=&2\int_0^1\int_0^{x_1}
   \frac{dx_1dx_2}{\left[[p_\parallel+(x_1v_\parallel + x_2u_\parallel)]^2-\Delta
   \right]^3},
   \label{Feynmanpar}
\eea
\end{widetext}
where $\Delta=x_1(x_1-1)v_\parallel^2 + x_2(x_2-1)u_\parallel^2 + 2x_2(x_1-1)(u\cdot v)_\parallel$.\\
For photons emitted at mid-rapidity the momentum components along the reaction plane are small. Since the reaction plane is perpendicular to the magnetic field we have $r_\perp=(u+v)_\perp\simeq 0$. The remaining component of the photon momentum is the one directed along the plane containing the magnetic field and we take this also as small, so that $r_3=(v+u)_3\simeq 0$. Also, because we  focus on describing the emission of real photons we have $r^2=(u+v)^2\simeq(u+v)_\parallel^2=0\simeq v_0^2+u_0^2 + 2(v\cdot u)_0\simeq v_0^2+u_0^2 + 2(v\cdot u)_\parallel$. The main thermal effect on low momentum gluons is their developing a thermal mass $m_g$. Therefore we can write $v_0^2=u_0^2\simeq m_g^2$ to find $u_\parallel^2\simeq v_\parallel^2\simeq m_g^2,
(v\cdot u)_\parallel \simeq -m_g^2, \Delta\simeq m_g^2(x_1-x_2)(x_1-x_2-1)$.
We  make the shift $p_\parallel\longrightarrow l_\parallel=p_\parallel+(x_1v_\parallel + x_2u_\parallel)$
and get rid of odd powers of $l_{\parallel\mu}$ in the numerator of the momentum integrand. The remaining terms can be computed using the well known relations
\bea
    \!\!\!\!\!\int\frac{d^dl}{(2\pi)^d}\frac{1}{[l^2-\Delta]^n}&=&\frac{(-1)^ni}{(4\pi)^{d/2}}
    \frac{\Gamma(m)}{\Gamma(n)}\left(\frac{1}{\Delta}\right)^{m}\nonumber\\
    \!\!\!\!\!\int\frac{d^dl}{(2\pi)^d}\frac{l_\mu l_\nu}{[l^2-\Delta]^n}&=&\frac{(-1)^{n-1}i}{(4\pi)^{d/2}}
    \frac{g_{\mu\nu}}{2}
    \frac{\Gamma(m')}{\Gamma(n)}\left(\frac{1}{\Delta}\right)^{m'}\!\!\!\!,
    \label{wellknown}
\eea
with $m=n-d/2$, $m'=n-d/2-1$, $d=2$, $n=3$. \\
To compute the integrals over the Feynman parameters $x_1$ and $x_2$ we use the principal value prescription. This results into a polynomial containing linear and cubic terms in the components $\alpha,\ \mu,\ \nu$ of $u_\parallel$ and $v_\parallel$.  
The final expression for the matrix element squared, after adding the contribution from the Feynman diagram in Fig.~\ref{fig1}b, summed over polarizations, becomes
\bea
   \sum_{\mbox{\small{pol}},f}\left| {\mathcal{M}} \right|^2&=&
   \left(\frac{6256}{2187}\right)
   \left(\frac{{\mathcal V}{\mathcal T}}{m_g^2}\right)\delta^4(r-v-u)\nonumber\\
   &\times&\sum_{f}\left(q_fg^2\right)^2\left(q_fB\right)^2e^{-2\frac{r_\perp^2}{3q_fB}},
   \label{sumfin}
\eea
where ${\mathcal V}{\mathcal T}$ represents the space-time volume of the reaction coming from squaring $(2\pi)^4\delta(r-v-u)$ and we have included the sum over the two light flavors $f=u,d$. The odd-looking factor $6256/2187\sim 2.86$ is obtained from the longitudinal piece of the matrix element squared after collecting the coefficients of the contraction of  the polynomial in the components of $u_\parallel$ and $v_\parallel$.
This remarkably simple result exhibits several interesting features. First, since the gluon thermal mass is $m_g\sim g T$, the photon emission probability is proportional to $g^2$ instead of $g^4$, i.e. it is not as suppressed as could be naively expected. Second, the emission probability is proportional to $B^2$ and contains a space volume factor ${\mathcal{V}}={\mathcal{A}}\times{\mathcal{L}}$, where ${\mathcal{A}}$ and ${\mathcal{L}}$ represent the transverse (with respect to the magnetic field) area and longitudinal length of the reaction zone, respectively. A factor ${\mathcal{T}}$  represents the time duration of the reaction. It has been shown that in a heavy-ion collision the product $B{\mathcal{T}}$ is approximately constant with energy~\cite{intensity2}. The flux factor $B{\mathcal{A}}$ also remains approximately constant, with varying impact parameter, given that the field strength is inversely proportional to the overlapping volume in the collision. Therefore the photon emission probability caused by the magnetic field can be expected to be only mildly dependent on the centrality of the reaction, and on the beam energy.\\ 
The invariant photon yield is obtained by integrating over the corresponding phase space weighed with the thermal distribution, namely
\bea
   \frac{r_0dN}{d^3r}&=&\frac{1}{2(2\pi)^3}\int\frac{d^3u}{2u_0(2\pi)^3}\int\frac{d^3v}{2v_0(2\pi)^3}\nn
   &\times&\sum_{\mbox{\small{pol}},f}\left| {\mathcal{M}} \right|^2
   n(u_0)n(v_0),
   \label{invyield}
\eea
where $n(E)=1/[\exp{\sqrt{E^2+m_g^2}/T}-1]$ is the Bose-Einstein distribution and $T$ is the temperature. Notice that the photons in the final state are not weighed by a thermal distribution since they basically escape from the interaction region once they are produced. The total number of photons $N$ is obtained by integrating the above yield over the photon momentum.
\begin{figure}
\begin{center}
\includegraphics[scale=0.4]{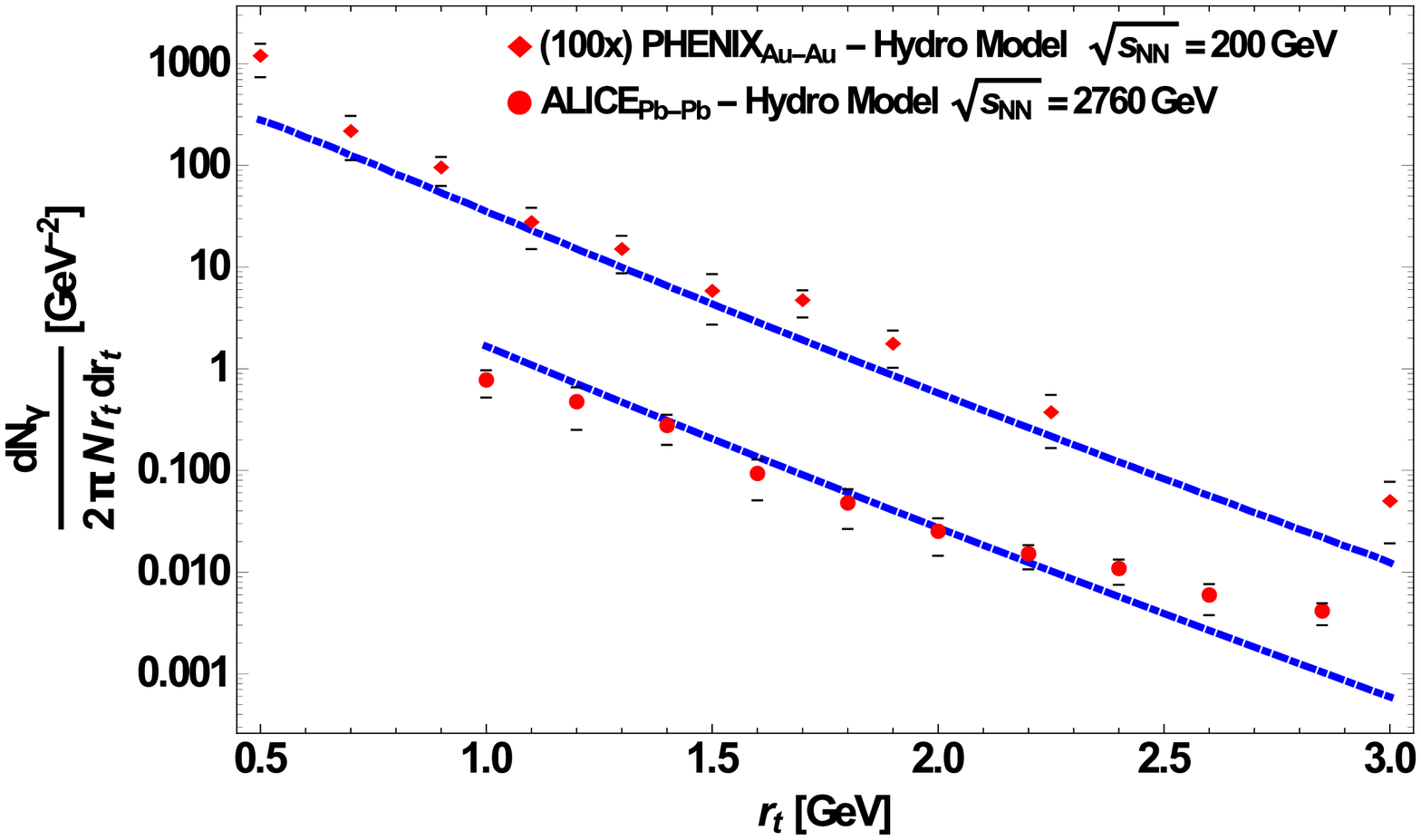}
\caption{Experimental excess photon yield with respect to the hydro calculation of Ref.~\cite{hydro-photons1} compared to our calculation for the centrality class $0-20\%$. The upper set corresponds to PHENIX data (multiplied by 100) and the lower set to ALICE data.}
\label{fig2}
\includegraphics[scale=0.4]{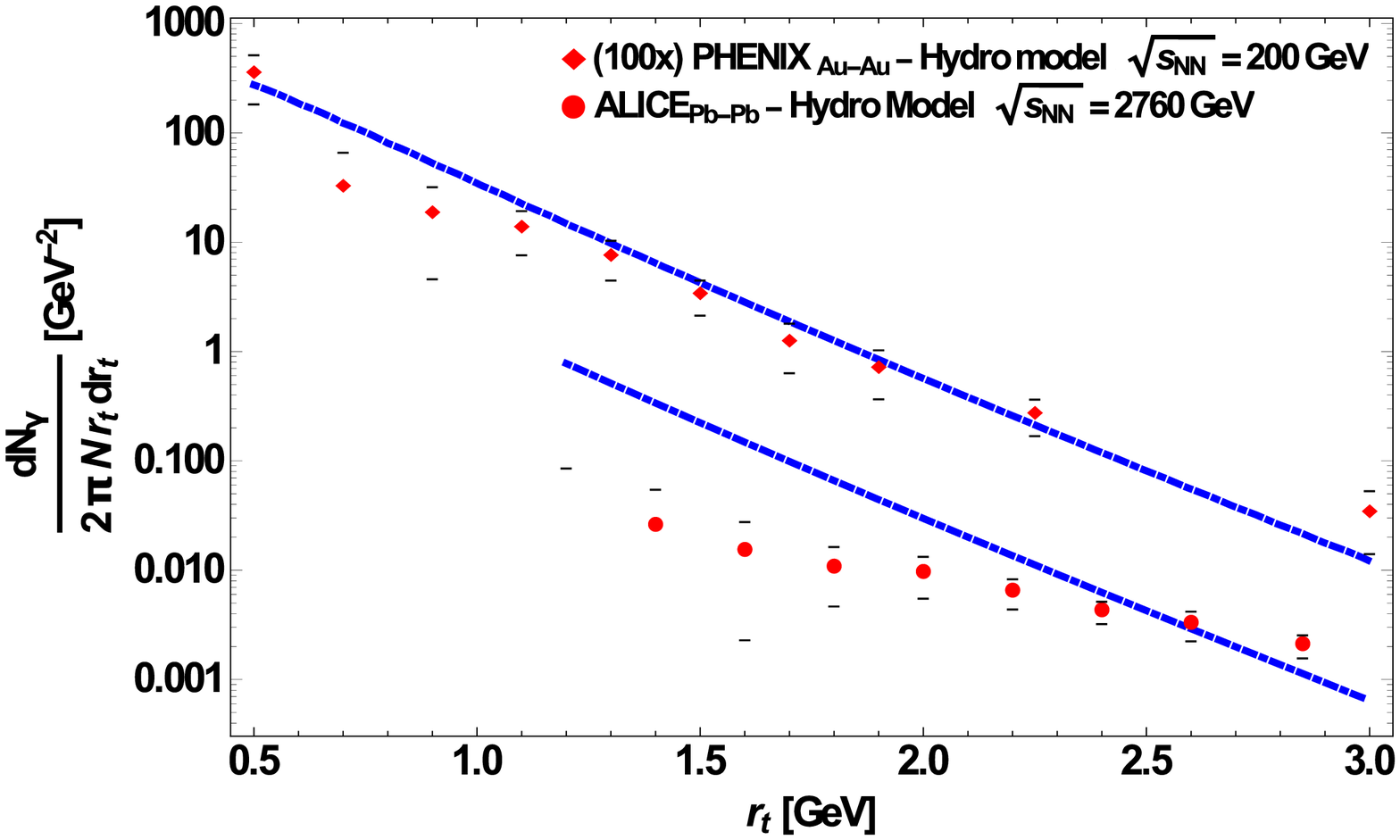}
\caption{{Experimental excess photon yield with respect to the hydro calculation of Ref.~\cite{hydro-photons1} compared to our calculation for the centrality class $20-40\%$. The upper set corresponds to PHENIX data (multiplied by 100) and the lower set to ALICE data.}}
\label{fig3}
\end{center}
\end{figure}

We write $d^3r/r_0=r_tdr_tdyd$ where $r_t$ is the magnitude of the photon momentum in the  plane transverse to the beam axis, $y$ the rapidity and $\phi$ the azimuthal angle. Notice that since $r_\perp=r_t\sinh (y)$, for $y\simeq 0$, $r_\perp\simeq yr_t$, $r_0=r_t\cosh (y)\simeq r_t$. Therefore, the number of photons per unit
momentum  transverse to the beam axis, integrated over the full azimuthal angle, and around mid-rapidity, is explicitly given by 
\bea
   \!\!\!\!\!\!\!\!\!\!\!\!\frac{dN}{dr_t}=
   {\mathcal{C}}
   \left[ \left(\frac{1}{3}\right)^{\! 4}\!\!\!e^{-2\frac{y_0^2r_t^2}{eB}} + 
   \left(\frac{2}{3}\right)^{\! 4}\!\!\!e^{-\frac{y_0^2r_t^2}{eB}}\right]\!\!I\left(\frac{r_t}{T};\sqrt{\frac{2}{3}g}\right)\!\!,
   \label{yield}
\eea 
where we have evaluated the distribution at $y=y_0=0.5$, given that the rapidity interval $\Delta y\sim 1$ is centered around mid-rapidity, and we have defined
\bea
   {\mathcal{C}}=\frac{g^4Te^4B^2}{8(2\pi)^7}
   \left(\frac{6256}{2187}\right)
   \left(\frac{{\mathcal V}{\mathcal T}}{m_g^2}\right),
   \label{defC}
\eea
and
\bea
  \!\!\!\!\!\!\!\!\!\! I(z;\lambda)\equiv\!\! \int_0^{z}\!\! \frac{dx x^2 n\left(\sqrt{(z+x)^2-(2x)^2}\right)n(x)}
   {\left(\sqrt{x^2+\lambda^2}\right)\left(\sqrt{(z+x)^2-(2x)^2+\lambda^2}\right)}.
   \label{defI}
\eea
We have also used that in QCD with two flavors, $m_g^2=(2/3)g^2T^2$. The yield given by Eq.~(\ref{yield}) can be properly called {\it thermal} because its computation assumes that the gluons are thermally distributed in phase space. However, notice that this yield {\it is not} proportional to $T^4$, as could be naively expected. Rather, it is inversely proportional to $T$, after accounting for the gluon mass squared in the denominator. This behavior is due to the fact that at the early stages of the collision it is the magnetic field which provides the largest of all energy scales. In fact, since we are working with the   energy scale hierarchy
\bea
   (eB)^{1/2} > T > m_g
   \label{hierarchy}
\eea
it is of no surprise that the yield is proportional to the fourth power of the largest energy involved, namely $(eB)^{1/2}$. The normalized yield is finally given by
\bea
   \frac{1}{2\pi Nr_t }\!\!\frac{dN}{dr_t}\!\!=\!\!\frac{\left[ \left(\frac{1}{3}\right)^4e^{-2\frac{y_0^2r_t^2}{eB}} + 
   \left(\frac{2}{3}\right)^4e^{-\frac{y_0^2r_t^2}{eB}}\right]\frac{I(r_t/T)}{2\pi}}
   {\sqrt{3\pi eB/2}\left[ \left(\frac{2}{3}\right)^{9/2} + \left(\frac{1}{3}\right)^{9/2}\right] 
   \int_0^\infty dr_tI(r_t/T)},
   \nonumber\\
   \label{nomalizedyield}
\eea
where the factor $\sqrt{3\pi eB/2}$ in the denominator comes from the integration over rapidity. \\
Notice that the normalized distribution is independent of the space-time region of the reaction. The impact parameter dependence is due to the dependence on the field intensity. The photon transverse yield given by Eq.~(\ref{nomalizedyield}) is an {\it excess yield} that should be added to calculations that do not consider magnetic field effects for photon emission.\\ 

In order to compare with experimental data we first proceed to use appropriate values for the temperature, the coupling $g$, and the magnetic field strength. Rather than pursuing an exhaustive search in the parameter space, we only consider here reasonable values for the above mentioned parameters.
We use $T=300$ MeV for RHIC and $T=350$ MeV for LHC. Also, since the analysis is valid for a gluon thermal mass smaller than $T$, we take $g=1$ (the results turn out to be only marginally sensitive to the value of $g$). The variation of the field intensity with time and impact parameter for RHIC and LHC energies is taken from Ref.~\cite{Zhong}. We chose one of the earliest times, $\tau\sim 0.05$ fm, hence one of the largest values of $eB$ which for RHIC, $\sqrt{s_{NN}}=200$ GeV, correspond to $0.5 \times 10^4 < eB/({\mbox{MeV}})^2 < 10^5$ and for LHC, $\sqrt{s_{NN}}=2.76$ TeV,  $eB/({\mbox{MeV}})^2\simeq 10^4$, with small variations coming from a slight dependence on the impact parameter. To relate the centrality class of the collision with the impact parameter we follow the geometrical model of Ref.~\cite{Broniowski}. Figures~\ref{fig2} and~\ref{fig3} show our results compared to the experimental excess photon yield with respect to one recent hydrodynamical calculation~\cite{hydro-photons1}.  The latter has been shown to approach the description of the experimental data within the lowest part of the uncertainties. Figure~\ref{fig2} (Fig.~\ref{fig3}) shows a comparison with the centrality class case $0-20\%$ ($20-40\%$). In each graph the upper set corresponds to PHENIX (multiplied by 100) and the lower to ALICE data. Notice that even with the ballpark choices of the parameters involved, our calculation provides a very good description of the excess photons. For the case of ALICE $20-40\%$ our calculation overshoots the data. This can be due to the fact that the hydrodynamic calculation~\cite{hydro-photons1} which we use as the reference to compute the excess also overshoots the data, for this centrality class at low $p_t$.\\

In conclusion our results support the idea that the excess in the thermal photon yield in heavy-ion reactions could be caused by the magnetic field induced emission of photons due to gluon fusion at the early stages of the reaction. The quantification of the asymmetry can also be pursued along these lines. This is work in progress that will be reported elsewhere.\\

The authors acknowledge useful conversations with M. E. Tejeda-Yeomans. Support for this work has been received in part from PAPIIT-UNAM under grant  number IN101515 and the National Research Foundation (South Africa), and the Harry Oppenheimer Memorial Trust OMT Ref. 20242/02.


\begin{thebibliography}{55}

\bibitem{RHIC}
I. Arsene {\it et al.} (BRAHMS Collaboration), Nucl. Phys. A757, 1 (2005); B. B. Back {\it et al.}, Nucl. Phys. A {\bf 757}, 28 (2005); J. Adams {\it et al.} (STAR Collaboration), Nucl. Phys. {\bf A757}, 102 (2005); K. Adcox {\it et al.} (PHENIX Collaboration), Nucl. Phys. {\bf A757}, 184 (2005); P. Jacobs and X. N. Wang, Prog. Part. Nucl. Phys. {\bf 54}, 443 (2005).

\bibitem{LHC}
F. Becattini, J. Phys. Conf. Ser. {\bf 527}, 012012 (2014) and references therein.

\bibitem{intensity1}
D. E.  Kharzeev,  L. D.  McLerran  and  H. J.  Warringa, Nucl. Phys. A {\bf 803}, 227 (2008).

\bibitem{intensity2}
V. Skokov, A.Y. Illarionov, V. Toneev, Int. J. Mod. Phys. A {\bf 24}, 5925 (2009).

\bibitem{intensity3}
V. Voronyuk, V. D. Toneev, W. Cassing, E. L. Bratkovskaya, V. P. Konchakovski, S. A. Voloshin, Phys. Rev. C {\bf 83}, 054911 (2011).

\bibitem{chargesep}
L. Adamczyk {\it et al.} (STAR Collaboration), Phys. Rev. Lett. {\bf 113}, 052302.

\bibitem{Fukushima}
K. Fukushima, Lect. Notes Phys. {\bf 871}, 241-259 (2013).

\bibitem{Skokov}
G. Basar, D. Kharzeev, V. Skokov, Phys. Rev. Lett. {\bf 109}, 202303 (2012).

\bibitem{Shuryak}
G. Basar, D.E. Kharzeev, E. V. Shuryak, Phys. Rev. C {\bf 90}, 014905 (2014).

\bibitem{Tuchin}
K. Tuchin, Phys. Rev. C {\bf 91}, 014902 (2015).

\bibitem{Leonardo}
G. Arciniega, P. Ortega, L. Pati\~no, J. High Energy Phys. {\bf 1404}, 192 (2014); S.~Y.~Wu and D.~L.~Yang, J. High Energy Phys. {\bf 1308}, 032 (2013) K.~A.~Mamo, J. High Energy Phys. {\bf 1308}, 083 (2013).

\bibitem{experiments}
A.~Adare {\it et al.} [PHENIX Collaboration], Phys.\ Rev.\ C {\bf 91}, 064904 (2015);   J.~Adam {\it et al.} 
[ALICE Collaboration], Phys.\ Lett.\ B {\bf 754}, 235 (2016).

\bibitem{McLerran}
L.~McLerran and B.~Schenke, Nucl. Phys. A {\bf 946}, 158 (2016).

\bibitem{Liu} 
F.~M.~Liu, S.~X.~Liu and K.~Werner, arXiv:1512.08833 [nucl-th].

\bibitem{v2photons}
A. Adare {\it et al.} [PHENIX Collaboration], e-Print: arXiv:1509.07758 [nucl-ex].

\bibitem{hydro-photons1}
J.-F. Paquet, C. Shen, G. S. Denicol, M. Luzum, B. Schenke, S. Jeon, C. Gale,  	e-Print:arXiv:1509.06738 [hep-ph].

\bibitem{hydro-photons2} 
H. van Hees, M. He, R. Rapp, Nucl. Phys. A {\bf 933}, 256Ð271 (2015).

\bibitem{transport}
O. Linnyk, V. Konchakovski, T. Steinert, W. Cassing, E. L. Bratkovskaya, Phys. Rev. C {\bf 92}, 054914 (2015).

\bibitem{review}
For a recent review see C. Shen, e-Print: arXiv:1601.02563 [nucl-th].

\bibitem{Schwinger} 
J. Schwinger, Phys. Rev. {\bf 82}, 664 (1951).

\bibitem{Zhong}
Y. Zhong, C.-B. Yang, X. Cai, S.-Q. Feng, Adv. High Energry Phys. {\bf 2014}, 193039 (2014).

\bibitem{Broniowski}
W. Broniowski, W. Florkowski, Phys. Rev. {\bf C} 65, 024905 (2002).

\end{thebibliography}
\end{document}